\documentclass[fleqn,usenatbib]{mnras}

\usepackage{newtxtext,newtxmath}

\usepackage[T1]{fontenc}

\DeclareRobustCommand{\VAN}[3]{#2}
\let\VANthebibliography\thebibliography
\def\thebibliography{\DeclareRobustCommand{\VAN}[3]{##3}\VANthebibliography}

\usepackage{graphicx}	
\usepackage{amsmath}	

\newcommand{\rbar}{$R_{\rm bar}$}

\newcommand{\omegabar}{$\Omega_{\rm bar}$}
\newcommand{\sbar}{$S_{\rm bar}$}

\newcommand{\rr}{${\cal R}$}
\newcommand{\vcirc}{$V_{\rm circ}$}

\title[Lopsided bar in IC~3167]{A slow lopsided bar in the interacting dwarf galaxy IC~3167}

\author[V. Cuomo et al.]{V. Cuomo,$^{1}$\thanks{E-mail: virginia.cuomo@uda.cl}
E. M. Corsini,$^{2,3}$
L. Morelli,$^{1}$
J. A. L. Aguerri,$^{4,5}$
Y. H. Lee,$^{6}$
L. Coccato,$^{7}$
A. Pizzella,$^{2,3}$
\newauthor{C. Buttitta,$^{2}$
and D. Gasparri$^{1}$}
\\
$^{1}$Instituto de Astronomía y Ciencias Planetarias, Universidad de Atacama, Avenida Copayapu 485, Copiapó, Atacama 1530000, Chile\\
$^{2}$Dipartimento di Fisica e Astronomia “G. Galilei”, Università di Padova, vicolo dell’Osservatorio 3, 35122 Padova, Italy\\
$^{3}$INAF – Osservatorio Astronomico di Padova, vicolo dell’Osservatorio 2, 35122 Padova, Italy\\
$^{4}$Instituto de Astrofísica de Canarias, calle Vía Lactea s/n, 38205 La Laguna, Tenerife, Spain\\
$^{5}$Departamento de Astrofísica, Universidad de La Laguna, Avenida Astrofísico Francisco Sánchez s/n, 38206 La Laguna,Tenerife, Spain\\
$^{6}$Korea Astronomy and Space Science Institute, 776 Daedeokdae-ro, Yuseong-gu 34055, Daejeon, Korea\\
$^{7}$European Southern Observatory, Karl-Schwarzschild-Strasse 2, D-85748 Garching, Germany}

\date{Accepted XXX. Received YYY; in original form ZZZ}

\pubyear{2022}

\begin{document}
\label{firstpage}
\pagerange{\pageref{firstpage}--\pageref{lastpage}}
\maketitle

\begin{abstract}
We present surface photometry and stellar kinematics of IC~3167, a dwarf galaxy hosting a lopsided weak bar and infalling into the Virgo cluster. We measured the bar radius and strength from broad-band imaging and bar pattern speed by applying the Tremaine-Weinberg method to stellar-absorption integral-field spectroscopy. We derived the ratio of the corotation radius to bar radius (${\cal{R}}=1.7^{+0.5}_{-0.3}$) from stellar kinematics and bar pattern speed.  The probability that the bar is rotating slowly is more than twice as likely as that the bar is fast. This allows us to infer that the formation of this bar was triggered by the ongoing interaction rather than to internal processes.
\end{abstract}




\section{Introduction} 
\label{sec:intro}

Barred galaxies are the most populated family of disc galaxies in the nearby Universe \citep[e.g.,][]{Aguerri2009, Buta2015} and their morphology, kinematics, and dynamics depend on the bar properties \citep[see][for a review]{Buta2013}. Indeed, bars reshape bulges, regulate star formation, and drive secular transformation of their host galaxies \citep[e.g.,][]{James2016, Lin2020}. Therefore, studying bars is a key task to understand disc galaxies and trace back their evolutionary pathway.

The main properties of a bar are the length \rbar, strength \sbar, and pattern speed \omegabar. In particular, \rbar\ measures the extension of the stellar orbits supporting the bar, \sbar\ quantifies the bar contribution to the galaxy gravitational potential, while \omegabar\ is the angular frequency of the bar figure rotation around the galaxy centre. This latter contributes to the bar rotation rate ${\cal R}=R_{\rm cor}/R_{\rm bar}$ where the corotation radius $R_{\rm cor} = V_{\rm circ}/\Omega_{\rm bar}$ and $V_{\rm circ}$ is the circular velocity. The value of ${\cal R}$ allows to distinguish between fast/long ($1.0<{\cal{R}}<1.4$) and slow/short (${\cal{R}}>1.4$) bars \citep{Athanassoula1992b,Debattista2000}. 

During the evolution of a barred galaxy, its \rbar\ and \sbar\ increase while \omegabar\ decreases since the bar exchanges angular momentum with the bulge, disc and dark matter (DM) halo. The dynamical friction induced by a centrally-concentrated DM halo efficiently brakes the bar and pushes \rr\ into the slow regime \citep[e.g.,][]{Debattista1998, Athanassoula2013, Petersen2019}. On the other hand, bars are expected to be born slow when their formation is triggered by gravitational interactions, as it is likely to occur in dense galaxy environments \citep[e.g.,][]{MartinezValpuesta2016, Lokas2018}.

In general, the measurement of \rbar\ and \sbar\ is based on image analysis and that of $V_{\rm circ}$ requires dynamical modelling, while  \omegabar\ can be derived through a variety of photometric, kinematic, and dynamical methods \citep[see][and references therein]{Rautiainen2008}. The only direct way to recover it was proposed by \citet[][hereafter TW]{Tremaine1984} and it is based on the straightforward equation $ \Omega_{\rm bar} \sin i = \langle V \rangle/ \langle X \rangle$, applied to a dynamical tracer satisfying the continuity equation, like the old stellar population in a dust-poor galaxy. In this case $\langle X \rangle$ and $\langle V \rangle$ are the luminosity-weighted averages of position and line-of-sight (LOS) velocity of the stars measured in apertures located parallel to the disc major axis and $i$ is the disc inclination. The TW method was widely applied to long-slit \citep[see][and references therein]{Corsini2011} and integral-field spectroscopic data \citep[see][and references therein]{Cuomo2020}. Observationally all the measured stellar bars resulted to be fast, suggesting that their formation was not triggered by an interaction and implying a low DM content in the central region of the host galaxies.

Stellar bars generally cross the galaxy centre and have a bisymmetric boxy shape \citep[e.g.,][]{Debattista2006, Mendezabreu2018}. However, some of them are asymmetric and off-centred with respect to the galaxy disc \citep[e.g.,][]{Odewahn1994, Kruk2017}. The most iconic example of a lopsided bar is hosted by the Large Magellanic Cloud \citep[LMC,][]{vanderMarel2001, JacyszynDobrzeniecka2016}. Off-centred bars are common in low luminosity galaxies with a companion \citep{Odewahn1994, Besla2016}. For this reason, their formation was considered as driven by tidal interactions \citep{Buta2001, Lokas2021b}. On the other hand, the presence of lopsided bars in non-interacting and isolated galaxies has been thought of as evidence for the gravitational pull of an asymmetric DM halo \citep{Kruk2017}. The effects of cosmological asymmetrical accretion of gas on galaxy discs can create strongly lopsided features, which should correspond to asymmetries in the star formation of the host galaxy \citep{Bournaud2005}.

Dwarf galaxies outnumber normal and giant galaxies \citep[e.g.,][]{McConnachie2012, Choque-Challapa2021} and are very common in dense environments, such as galaxy groups and clusters, which play an important role in shaping their morphology and stellar properties \citep[see][for a review]{Boselli2014}. Some dwarf disc galaxies have bars, lenses, and spiral arms as their giant counterparts and are characterised by a strong asymmetric shape \citep[e.g.,][]{Barazza2002, Lisker2006, Michea2021}. We consider dwarf barred galaxies as the ideal candidates to host slow-rotating bars because dwarf galaxies are thought to be embedded in massive and quite centrally-concentrated DM halos \citep[e.g.,][]{Adams2014, Relatores2019},  which may dynamically brake the bar and are prone to interactions with other galaxies and/or cluster triggering the formation of a slow bar. 

\begin{figure}
    \centering
    \includegraphics[scale=0.35]{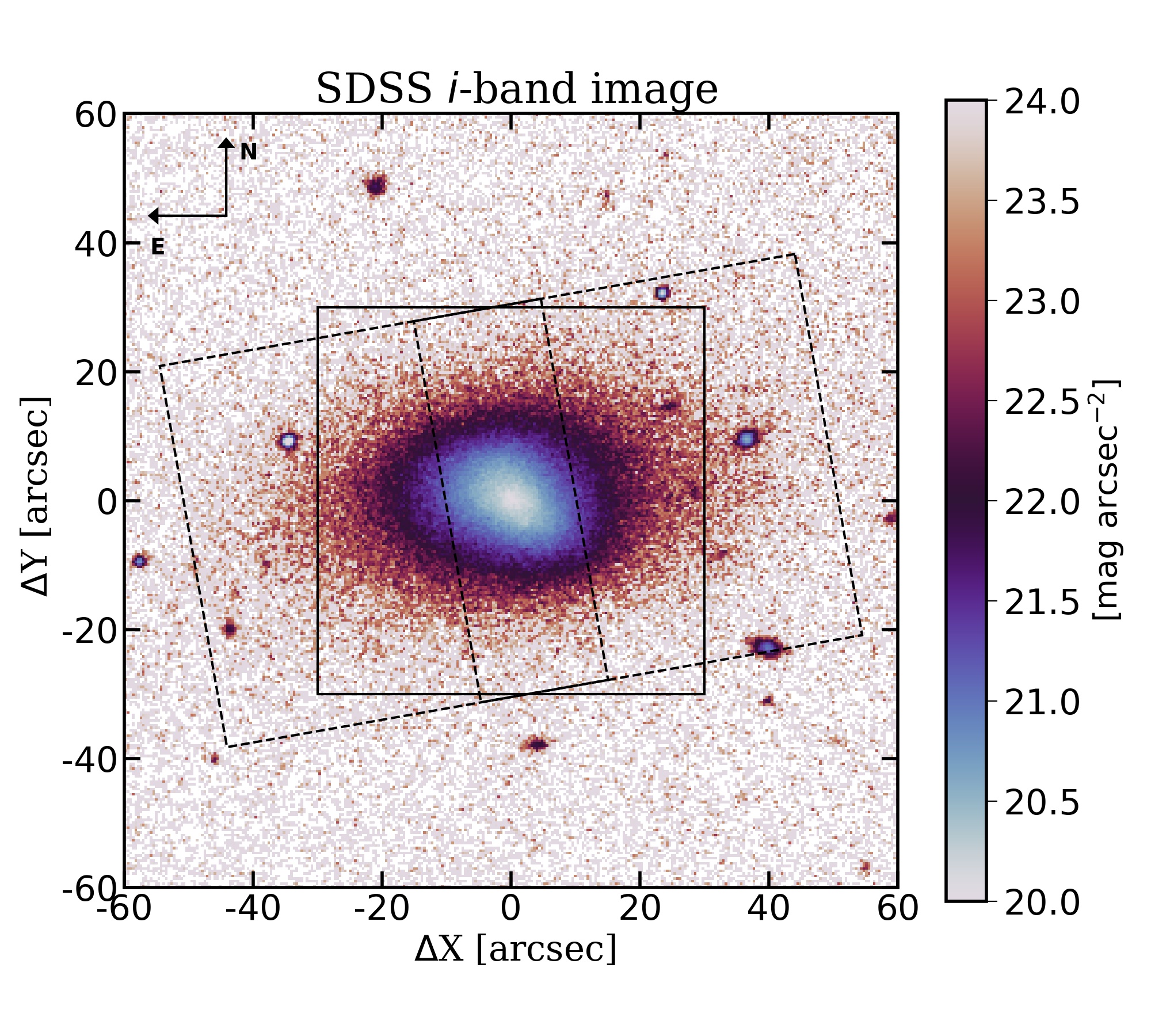}
    \caption{SDSS $i$-band image of IC~3167. The bar and the disc extend up to $\sim15$ and $\sim50$ arcsec, respectively. The black contours show the field of view of the MUSE pointings.}
    \label{fig:img_sdss}
\end{figure}

However, \rr\ is poorly known in dwarf galaxies, due to the difficulty of accurately measuring their \omegabar\ \citep{Corsini2007}. To start addressing this issue, here we report a detailed photometric and kinematic study of IC~3167 (VCC~407, Fig.~\ref{fig:img_sdss}). 
This is a dwarf lenticular galaxy in the Virgo cluster \citep{Kim2014} with $r$-band effective radius $R_{{\rm e}, r} = 1.5$ kpc and absolute magnitude $M_r=-17.62$ mag \citep{Lisker2006} assuming a distance of 17 Mpc. Its stellar mass is $M_\ast = 10^{9.06}~{\rm M}_{\odot}$ \citep{Bidaran2020}. 
In spite of being initially classified as an early-type dwarf galaxy, IC~3167 hosts an inclined disc \citep{Lisker2006} and a bar ($R_{\rm bar}\sim14.0$ arcsec, \citealt{Janz2014}).
The galaxy does not have any bright nearby companion, but it belongs to a bound group of dwarf galaxies recently accreted onto the Virgo cluster and observed at a cluster-centric distance of $\sim1.5$ Mpc with a LOS velocity of $\sim700$ km~s$^{-1}$ with respect to M87 \citep{Lisker2018}.
%

\begin{figure*}
    \centering
    \includegraphics[scale=0.51]{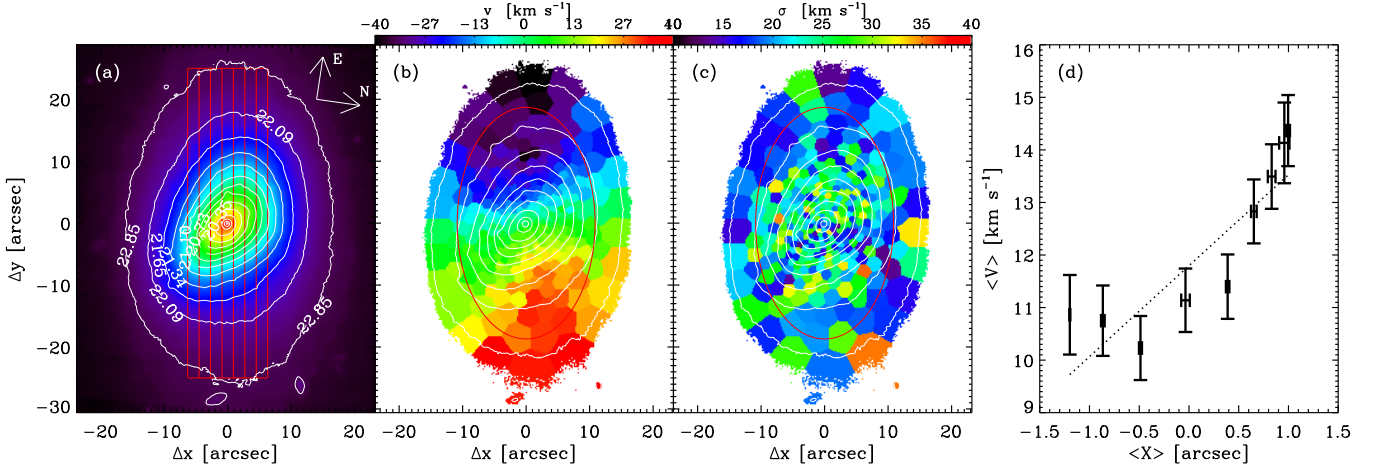}
    \caption{Stellar kinematics and bar pattern speed of IC~3167. {\em Panel (a)\/}: MUSE reconstructed image. The white lines mark a few isophotes to highlight the orientation of the lopsided bar and disc. The surface brightness level is reported for the outer isophotes and for every three internal ones, after calibrating the MUSE reconstructed image to the SDSS $i$-band. The red rectangles show the location of the pseudo-slits adopted to derive \omegabar. {\em Panel (b)\/}: Map of the LOS velocity subtracted of systemic velocity. The value of \vcirc\ was derived from spatial bins outside the red ellipse. {\em Panel (c)\/}: Map of the LOS velocity dispersion corrected for instrumental velocity dispersion. {\em Panel (d)\/}: Kinematic integrals $\langle V \rangle$ plotted as a function of photometric integrals $\langle X \rangle$. The best-fitting straight dotted line has a slope $\Omega_{\rm bar}\sin i= 2.01\pm 0.38$ km s$^{-1}$ arcsec$^{-1}$.}
    \label{fig:muse}
\end{figure*}

\section{Observations and data reduction}
\label{sec:observations}

The integral-field spectroscopic observations of IC~3167 were carried out with the Multi-Unit Spectroscopic Explorer (MUSE) of the European Southern Observatory (Fig.~\ref{fig:muse}). MUSE was configured in wide field mode to ensure a field of view (FOV) of $1\times1$ arcmin$^2$ with a spatial sampling of $0.2$ arcsec pixel$^{-1}$ and to cover the wavelength range of $4800-9300$ \AA\ with a spectral sampling of $1.25$ \AA\ pixel$^{-1}$ and an average nominal spectral resolution of ${\rm FWHM} = 2.51$ \AA\ \citep{Bacon2010}. The central pointing was obtained on January 2017 and February 2018 (120 min; Prog. Id. 098.B-0619(A) and 0100.B-0573(A), P.I.: T. Lisker). On April 2021, we took two offset pointings along the galaxy major axis at a distance of 20 arcsec eastward (20 min) and westward (10 min) from the galaxy nucleus (Prog. Id.: 0106.B-0158(A), P.I.: V. Cuomo). During the nights the seeing reached a mean value of ${\rm FWHM_{seeing}}\sim 1.1$ arcsec.
We performed the data reduction as detailed in \citet{Cuomo2019a}, using the standard MUSE pipeline \citep[version 2.8.4,][]{Weilbacher2020}, including bias and overscan subtraction, flat fielding, wavelength calibration, determination of the line spread function, sky subtraction, and flux calibration. The sky contribution was quantified using an on-sky exposure. Then, we determined the effective spectral resolution and its variation across the FOV, and produced the combined datacube of the galaxy.

Moreover, we retrieved the flux-calibrated $i$-band image (54 sec) of IC~3167 from the science archive of the Sloan Digital Sky Survey Data Release 14 \citep[SDSS,][]{Abolfathi2018}. We fitted ellipses to the galaxy isophotes and beyond among with the \textsc{iraf} task \textsc{ellipse} \citep{Jedrzejewski1987}, to measure the constant residual surface brightness of the sky, which we subtracted as done in \cite{Morelli2016}.

\section{Data analysis and results} 
\label{sec:analysis}

\subsection{Isophotal and Fourier analysis} 
\label{sec:photometry}

\begin{figure*}
    \centering
    \includegraphics[scale=0.47]{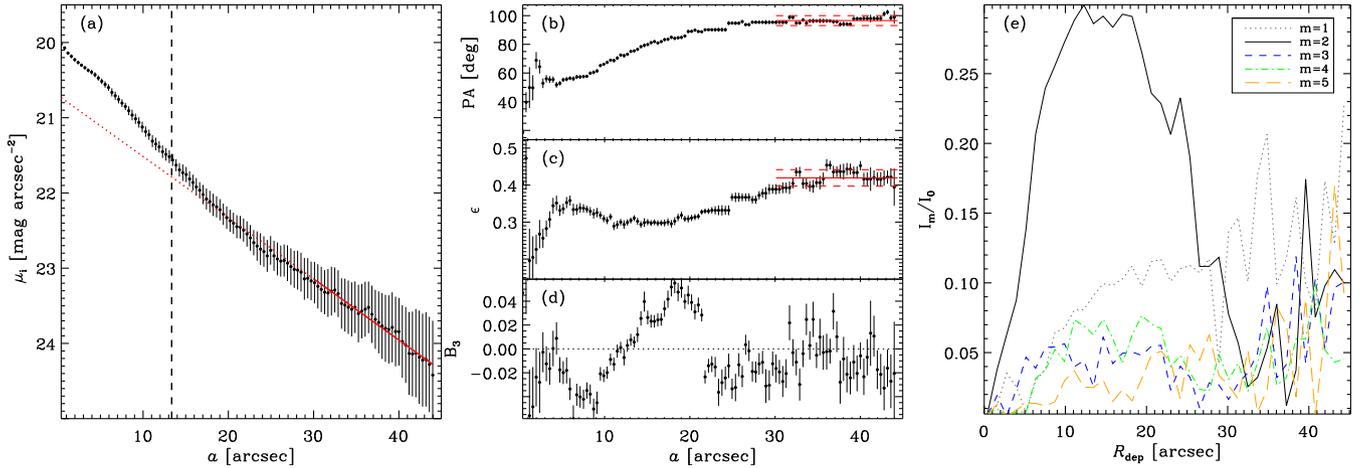}
    \caption{Isophotal and Fourier analysis of the $i$-band image of IC~3167. {\em Panels (a)-(d)\/}: Radial profiles of azimuthally-averaged surface brightness, position angle, ellipticity, and third sine Fourier coefficient. The red line in panel (a) marks the best-fitting disc surface brightness with the solid portion corresponding to the region used to fit the disc exponential profile. The vertical black dashed line marks the disc scale-length. The horizontal red solid and dashed lines in the panels (b) and (c) give the mean values and rms of PA and $\epsilon$ measured for the disc. {\em Panel (e)\/}: Smoothed radial profiles of the relative amplitude of the $m=0,1,2,3,4$ and 5 Fourier components.}
    \label{fig:sdss}
\end{figure*}

From the isophotal analysis of the $i$-band image of IC~3167, we derived the radial profiles of the azimuthally-averaged surface brightness, ellipticity $\epsilon$, position angle PA, centre coordinates and third, fourth, and sixth cosine and sine Fourier coefficients describing the deviation of the isophotal shape from a perfect ellipse out to $\sim50$ arcsec from the centre, where the surface-brightness level of the sky was reached. 
The local maximum of ellipticity $\epsilon\sim0.35$ associated to the nearly constant position angle ${\rm PA}\sim55^\circ$ in the inner $\sim10$ arcsec is the isophotal signature of the bar, which is offset southward by $0.7$ arcsec with respect to the disc centre. Moreover, the bar is remarkably lopsided, as it results from the third sine Fourier coefficient peaking at $B_3\sim-0.05$. The inner portion of the disc is lopsided too ($B_3\sim0.06$, Fig.~\ref{fig:sdss}). 
We measured the disc ${\rm PA}=96.5\degr \pm 1.4\degr$ and $\epsilon=0.419\pm0.016$ from the constant isophotal profiles between 30 and 45 arcsec. This latter value was adopted to recover the galaxy inclination as $ i=acos(1-\epsilon)=54.5\degr \pm 1.1\degr$ assuming an infinitesimally-thin disc, following \citet{Cuomo2019a}.

We obtained consistent results from the Fourier analysis of the $i$-band image, which we deprojected by adopting the disc geometric parameters. We derived the radial profiles of the amplitude of the $m=0,1,2,3,4$ and 5 Fourier components and phase angle of the $m=2$ one as in \citet{Aguerri2000}.
The large values of the $m=2$ component with a maximum $I_2/I_0\sim0.3$ and its constant phase angle $\phi_2 \sim 40\degr$ are the Fourier signatures of the bar. However, the asymmetric radial profile of the $m=2,4$ components shows that the bar does not have a bisymmetric shape. Moreover, the $m=1,3,5$ components have large values within the region of the bar confirming its lopsidedness (Fig.~\ref{fig:sdss}).

\subsection{Bar length and strength} 
\label{sec:length}

\begin{figure}
    \centering
    \includegraphics[scale=0.56]{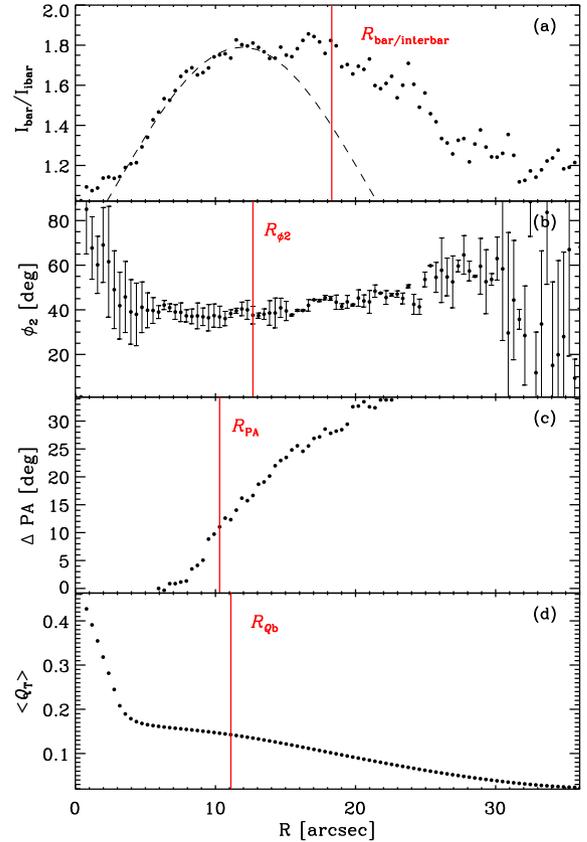}
    \vspace{-0.2cm}
    \caption{Bar length estimates from the $i$-band image of IC~3167. {\em Panels (a)-(d)\/}: radial profiles of the bar/interbar intensity ratio, phase angle, difference between PA of the isophotes in the region of the $\epsilon$ peak and of the bar, and transverse-to-radial force ratio. In each panel the vertical red line marks the corresponding \rbar.}
    \label{fig:rbar}
\end{figure}

We measured \rbar\ by analysing the $i$-band image of IC~3167 with four independent methods based on the bar/interbar intensity ratio (\citealt{Aguerri2000}; $R_{\rm bar/interbar} = 18.3$ arcsec), constant phase angle of the $m=2$ Fourier component (\citealt{Debattista2002}; $R_{\phi_2}=15.4$ arcsec), constant PA of the isophotes in the region of the $\epsilon$ peak (\citealt{Aguerri2009}; $R_{\rm PA} = 10.2$ arcsec), and location of a peak/plateau in the azimuthally-averaged radial profile of the transverse-to-radial force ratio (\citealt{Lee2020}; $R_{Q_{\rm b}} = 11.1$ arcsec). 
We adopted the mean value of four measurements (Fig.~\ref{fig:rbar}) to get $R_{\rm bar}=13.1^{+5.2}_{-2.8}$ arcsec, where the upper/lower error is the largest deviation of the highest/lowest estimate from the mean. This value is consistent within the errors with previous results based on photometric decomposition by \citet{Janz2014}.

We measured \sbar\ from the maximum $I_2/I_0$ ratio between the amplitudes of $m = 2$ and $m = 0$ Fourier components (\citealt{Athanassoula2002}; $S_{\rm Fourier} = 0.31$) and from the transverse-to-radial force ratio map (\citealt{Lee2020}; $S_{Q_{\rm b}}=0.23$). This gives $S_{\rm bar} = 0.27\pm0.04$, which means that the bar of IC~3167 is weak according to the classification by \citet{Cuomo2019b}.

\subsection{Stellar kinematics and circular velocity}
\label{sec:kinematics}

We measured the LOS stellar velocity and velocity dispersion of IC~3167 from the MUSE combined datacube using the Galaxy IFU Spectroscopy Tool pipeline \citep[GIST,][]{Bittner2019}. As done in \citet{Cuomo2019a}, we performed a Voronoi binning with a target signal-to-noise ratio of 60 per bin, adopted the MILES stellar library \citep{Vazdekis2010} in the wavelength range $4800-5600$ \AA, which is the same we adopted to apply the TW method. We estimated the errors on the kinematic parameters using Monte Carlo simulations. Our measurements (Fig.~\ref{fig:muse}) are in agreement with those obtained by \citet{Bidaran2020} for the central pointing.

We derived the circular velocity $V_{\rm circ}=53.8\pm1.5$ km s$^{-1}$ from the LOS stellar velocity and velocity dispersion measured in the disc spatial bins outside the bar dominated region (i.e., outside the ellipse with semi-major axis of $18.7$ arcsec) applying the asymmetric drift correction by \cite{Binney2008} and following the prescriptions of \citet[][Fig.~\ref{fig:muse}]{Aguerri2003,Aguerri2015,Cuomo2019a}. Moreover, the disc scale-length ($h=13.4\pm1.1$ arcsec) was derived by fitting an exponential law to the surface-brightness radial profile in the disc-dominated region (i.e. for radii larger than $30$ arcsec, Fig.~\ref{fig:sdss}).

\subsection{Bar pattern speed and rotation rate} 
\label{sec:tw}

We applied the TW method to the MUSE combined datacube of IC~3167 to measure \omegabar\ for its bar. To this aim, we defined 7 adjacent pseudo-slits aligned with the disc PA and crossing the bar (Fig.~\ref{fig:muse}). They have a width of 9 pixels (1.8 arcsec, i.e., slightly larger than ${\rm FWHM_{seeing}}$) to deal with seeing smearing effects and a half-length of 125 pixels (25 arcsec) to reach the disc and cover the radial region where the integrals converge to a constant value \citep{Zou2019}.

We measured the photometric integrals from the reconstructed image of IC~3167, which we obtained by summing the MUSE combined datacube in the wavelength range $4800-5600$ \AA, which is well suited for the application of the TW method as discussed by \cite{Cuomo2019a}. In each pseudo-slit we derived the radial profile of the total surface-brightness and calculated the luminosity-weighted distance $\langle X \rangle$ of the stars from the galaxy minor axis. 
We measured the kinematic integrals from the MUSE combined datacube in the wavelength range $4800-5600$ \AA . In each pseudo-slit we collapsed all the spaxels into a single spectrum and derived the luminosity-weighted LOS velocity $\langle V \rangle$ of the stars.
For each pseudo-slit we estimated the errors associated to $\langle X \rangle$ by calculating the root mean square of the values of the integrals obtained varying the slit length in the convergence region \citep{Zou2019} and to $\langle V \rangle$ running Monte Carlo simulations on a set of mock spectra, respectively.


A linear correlation is expected for a bar tumbling as a rigid body \citep{Corsini2003, Meidt2008}. We derived $\Omega_{\rm bar}\sin i= 2.01\pm 0.38$ km s$^{-1}$ arcsec$^{-1}$ by fitting with a straight line to the $\langle X \rangle$ and $\langle V \rangle$ values and their errors using the \textsc{IDL FITEXY} algorithm (Fig.~\ref{fig:muse}). This corresponds to $\Omega_{\rm bar} = 2.47\pm0.45$ km s$^{-1}$ arcsec$^{-1}$, which is $30.0\pm5.4$ km s$^{-1}$ kpc$^{-1}$.

We obtained ${\cal R}=1.7^{+0.5}_{-0.3}$ for the adopted values of $i$, \rbar, \vcirc, and \omegabar\ and performing a Monte Carlo simulation to account for their errors. The resulting value of \rr\ lies just above the limit for a bar to be fast. The lopsided bar of IC~3167 is more than twice more likely to be slow (probability of 68\%) rather than fast (32\%), after excluding the ultrafast regime (2\%). We refer to \cite{Cuomo2021} for a discussion of this latter unphysical case.

\section{Discussion and conclusions} 
\label{sec:conclusions}

In this paper we present a photometric and kinematic analysis of the lopsided stellar bar hosted in the dwarf lenticular galaxy IC~3167 located in the Virgo cluster. 

The galaxy shows an elongated off-centred stellar structure with an uncommon triangular shape embedded in the galaxy disc (see Figs.~\ref{fig:img_sdss} and \ref{fig:muse}). 
We confirmed that this peculiar structure is a genuine bar by means of archival $i$-band SDSS imaging and customised MUSE integral-field spectroscopy.
In particular, we observed:\\
- ellipticity and position angle radial profiles typical for a barred galaxy;\\
- strong and peaked profile of the Fourier $m=2$ component;\\
- four thick slabs from the corners of the bar in the map of the transverse-to-radial force ratio.\\
%
The triangular shape of the bar leads to the large values of the third sine Fourier coefficient of the isophotal analysis and $m=3,5$ components of the Fourier analysis.
Moreover, the application of the TW method to the spectroscopic data gave the pattern speed of the bar. It is tumbling as a rigid body with a rotation rate ${\cal{R}}=1.7^{+0.5}_{-0.3}$ consistent with the slow bar regime. 

Off-centred bars are common in low-luminosity Magellanic-type galaxies \citep{Odewahn1994, Kruk2017}, while lopsided bars are more rare: the LMC hosts the first convincing example of an asymmetric stellar bar \citep{vanderMarel2001} being as well slightly off-centred with respect to the disc \citep{JacyszynDobrzeniecka2016}. Another recently found example is the dwarf irregular galaxy DDO~168, which hosts a lopsided gaseous bar \citep{Patra2019}. 

Bars are typically bisymmetric and elongated structures supported by stars moving in symmetric and elongated periodic orbits belonging to the so-called $x_1$ family and to the vertically-extended families bifurcating from it \citep{Skokos2002}.  Nevertheless, stable regular and stochastic orbits with asymmetric morphologies can also be present \citep{Athanassoula1992a, Voglis2007}. They could be the backbone of the orbital structure of lopsided bars although it is not yet clear which internal or external process activate this kind of orbits.

Both off-centred and asymmetric bars have been observed in simulated galaxies as the end result of different formation and evolution scenarios. 
An off-centred bar, one-arm spiral and one-sided star formation can be induced by a short tidal interaction \citep{Yozin2014}. These asymmetries can widely vary in amplitude and can be both short-lived or more persistent, especially in low-mass galaxies. They can occur at the bar formation or later in its evolution. Recently, the presence of lopsided bars has been reported in the Illustris TNG100 simulation \citep{Lokas2021b}. These asymmetric bars are found in a small fraction ($\sim5\%$) of barred-like galaxies, where the bulk of the stars typically forms an elongated structure with a little amount of gas \citep{Lokas2021a}. 

\cite{Lokas2021b} investigated two scenarios leading to the formation of a lopsided bar using numerical simulations of galaxy evolution in the cosmological context. First, they considered the interaction between a Milky Way-like barred galaxy and a massive satellite, which is moving onto a radial orbit in the disc plane and perpendicular to the bar at the time of the first flyby. In addition, they analysed the secular evolution of a disc galaxy off-centred with respect to its DM halo. The bars formed in such simulations show some degree of displacement and asymmetry, as it results from their isophotal and Fourier analysis.
When the lopsidedness is driven by interaction, the forming bar survives and it becomes stronger and lopsided, because of the asymmetry in the effects of the satellite flybies on the two bar sides. In this case, the $m=3,5$ Fourier components present large values within the bar region, while the $m=1$ component gradually increases in the disc. On the contrary, a lopsided bar formed in an off-centred disc is characterised by smaller values of the $m=3,5$ Fourier components. The $m=1$ component is initially strong for the disc, but then it  decreases during the bar formation. 
The photometric properties of the bar of IC~3167 are consistent with a formation scenario driven by an interaction. Indeed, we measured large odd Fourier components within the bar region and a significant increase of the $m=1$ component in the disc region.

IC~3167 is member of a small group of early-type dwarf galaxies in the initial phase of accretion onto the Virgo Cluster \citep{Lisker2018}. \cite{Bidaran2020} measured the stellar kinematics to derive the specific angular momentum $\lambda_R$ of IC~3167 to explore the role of the environment in transforming late-type star-forming galaxies into quiescent spheroids. IC~3167 has a steep $\lambda_R$ radial profile and is a fast-rotating galaxy, which means that both the Virgo environment and processing mechanisms occurred in the host halo before the infall started $\sim2$ Gyr ago have been marginal so far. Nevertheless, this does not exclude that the formation of the bar could be triggered by flybies with other galaxies. 
Fast interactions are indeed predicted to not strongly affect the kinematical properties of the galaxy, except for a small increase of the velocity dispersion in the outer part of the disc. Moreover, bars induced by fast interactions are born slow and stay slow during their evolution. Finally, they are weaker than bars formed by internal disc instabilities \citep{MartinezValpuesta2016, Lokas2018}. 
Again, our observational findings of a weak and slowly-rotating bar in IC~3167 further support a formation induced by an ongoing interaction within the Virgo cluster.


Slow bars are also expected to be the result of an efficient dynamical friction exerted by the DM halo, a phenomenon which should be particularly efficient when a large amount of DM is present within the central part of the galaxy as expected for dwarf objects like IC~3167 \citep{Debattista2000, Sellwood2008, Fragkoudi2021}. 
Despite the presence of a massive and centrally-concentrated DM halo may have efficiently slowed down the rotation of the bar of IC~3167, its peculiar shape and rotation regime are consistent with a formation scenario driven by interaction. 

Galaxies hosting a lopsided bar are quite rare and remain a poorly known class of objects. In fact, IC~3167 is only the third galaxy in which the photometric and kinematic properties of its asymmetric and off-centred bar have been studied in detail. 

\section*{Acknowledgements}

We are grateful to the anonymous referee for the constructive suggestions which helped us to improve the manuscript. We thank A. Boselli, V. P. Debattista, L. Ferrarese, and S. Zarattini for their comments. 
VC is supported by the Chilean Fondecyt Postdoctoral programme 2022 No. 3220206 and by the ESO-Government of Chile Joint Committee programme ORP060/19 and thanks Instituto de Astrofísica de Canarias for hospitality during the preparation of this paper. 
CB, EMC, and AP are supported by MIUR grant PRIN 2017 20173ML3WW-001 and Padua University grants DOR2019-2021. 
JALA is supported by the Spanish Ministerio de Ciencia e Innovacion by the grant PID2020-119342GB-I00.

\section*{Data availability}

The derived data in this article will be shared on request to VC.




\bibliographystyle{mnras}




\bsp	
\label{lastpage}
\end{document}